\documentclass[prd,aps,nofootinbib,onecolumn,showpacs,12pt]{revtex4-1}
\usepackage{graphicx,epsfig}

\usepackage{epsf}
\usepackage{graphicx}

\begin{document}

\title{  \bf  On the mass and the decay constant of recently observed bound state $h_b(1P)$}
\author{ V. Bashiry$^{\dag1}$ \\
$^{\dag}$Engineering Faculty, Cyprus International University, Via Mersin 10, Turkey\\
$^1$bashiry@ciu.edu.tr}

\begin{abstract}
The mass and decay constant of the ground state of spin singlet  $b \bar{b}$ is calculated by using the QCD sum rules method. This meson is  an  axial vector $P$-wave  named $h_b(1P)$ meson with $J^P  =1^+$. It is found that the mass of this meson is $(m=9940\pm 37)$ MeV, which is  consistent with the recent experimental data by Belle collaboration. 
\end{abstract}

\maketitle

\section{Introduction}
Heavy quarkonium bound state and their  spectroscopy such as $b\bar{b}$ and $c\bar{c}$
have been considered as a good  tool for the investigation of the Quantum  Chromodynamics(QCD) interaction.  Bottomonium is specially important  because of the variety of bound states and decay channels allowing us to study and fix many parameters and aspects of Standard Model(SM)  and QCD.   Moreover, measurements of bottomonium masses, total widths
and transition rates serve as important benchmarks for
the predictions of QCD-inspired potential models, non-relativistic
QCD, lattice QCD and QCD sum rules \cite{Adachi:2011ji,Brambilla}. The
spin-singlet $P$-wave bound states of $b\bar{b}$ such as  $h_b(1P)$ can be used 
as tests of the $P$-wave spin-spin (or hyperfine) interaction. Therefore, theoretical calculations on the physical parameters of this meson and
their comparison with experimental data could give essential information about their nature and  hyperfine interaction.

The QCD sum rules approach \cite{svz,colangelo,braun,balitsky,Kazem1, Bashiry:2011pp},  which enjoys two peculiar properties namely its foundation based on QCD Lagrangian and free of model dependent parameters,  as one of the most well established  non-perturbative methods has widely been applied to hadron physics.
In the present work,  we extend the application of this method to calculate the masses and decay constants of  the axial vector $P$-wave  $h_b(1P)$ meson with the quantum numbers $J^P  =1^+$. Since the heavy quark
condensates are suppressed by the inverse powers of the heavy quark mass \cite{Narison, Jamin:1997rt}, so as the first nonperturbative contributions, we take into account and calculate the two-gluon condensate diagrams.  Note that, Belle collaboration has recently  reported the observation of the $h_b(1P)$  spin-singlet bottomonium state produced in the reaction $e^+e^- \rightarrow  h_b(1P)\pi^+\pi^−$ with significances of 5.5$\sigma$ . They found that
$m[h_b(1P)] = (9898.25\pm1.06^{+1.03}_{-1.07} )$ MeV \cite{Adachi:2011ji}. 

The layout of the paper is as follows: in the next section, the sum rules for the mass and decay constant of the ground state axial vector meson is calculated. Section III encompasses our numerical predictions for the mass and leptonic decay constant of the $h_b(1P) $ axial vector meson.

\section{QCD sum rules for the mass and decay constant of the $h_b(1P) $ axial vector meson }
The two point correlation function respunsible for the mass and decay constant of the axial vector meson can be written as:
\begin{eqnarray}\label{correl.func.}
\Pi^{A}_{\mu\nu} =i\int d^{4}xe^{ip. x}{\langle} 0\mid{\cal T}\left ( J_\mu^{A} (x) \bar J_\nu^{A}(0)\right)\mid0{\rangle},
\end{eqnarray}
where  ${\cal T}$  is the time ordering product and $J_\mu^{A}=\overline{b}(x)\gamma_\mu\gamma_5 b(x)$ is responsible for creating the  axial vector quarkonia from the vacuum with the same quantum numbers as the interpolating currents.

According to  the QCD sum rules, the correlation function has to be calculated in two different ways. Regarding the physical or phenomenological side, it can be evaluated in terms of hadronic parameters
such as meson mass and decay constant. On the other hand, in the QCD or theoretical side, it is calculated in terms of QCD degrees of freedom such as quark masses and gluon condensates by means  of the operator product expansion (OPE) in deep Euclidean region. Matching  these two representations of the correlation function via the dispersion relation, the relation among the physical or hadronic  parameters i.e., the mass and the decay constant and the fundamental QCD parameters can be achieved. The Borel transformation with respect to the momentum squared is applied  for both sides of the correlation function. This end is necessary to suppress the contributions of higher states and continuum.

Firstly, to calculate the physical part  a complete
set of intermediate  states with the same quantum numbers as the interpolating  current is added. Performing the integral over $x$ and isolating the ground state it is  obtained that:
\begin{eqnarray}\label{phen1}
\Pi^{A}_{\mu\nu}=\frac{{\langle}0\mid J^{A}_\mu(0) \mid A\rangle \langle A\mid J^{A}_\nu(0)\mid
 0\rangle}{m_{A}^2-p^2}
&+& \cdots,
\end{eqnarray}
 where $\cdots$ represents  contribution of the higher states and continuum and $m_{A}$ is  the mass of the axial vector meson.

 The matrix element, parametrized in terms of the leptonic decay constant, is as follows:
\begin{eqnarray}\label{lep}
\langle 0 \mid J(0)\mid A\rangle&=&f_{A}   m_{A} \varepsilon_\mu,
\end{eqnarray}
where $f_{A}$ is the leptonic decay constant of the axial vector meson.
The summation over the polarization vectors are:
\begin{eqnarray}\label{polvec}
\epsilon_\mu\epsilon^{*}_\nu&=&-g_{\mu\nu}+\frac{p_{\mu}p_{\nu}}{m_{A}^2},\nonumber\\
\end{eqnarray}
Using  Eq. (\ref{polvec}), the final expressions of the physical side of the correlation function is as:
\begin{eqnarray}\label{phen2}
\Pi^{A}_{\mu\nu}&=&\frac{f_{A}^2 m_{A}^2}{m_{A}^2-p^2}\left[-g_{\mu\nu}+\frac{p_{\mu}p_{\nu}}{m_{A}^2}\right] + \cdots,\nonumber\\
\end{eqnarray}
to proceed with the calculation of the mass and the decay constant the  $g_{\mu\nu}$ structure is used.

\begin{figure}[h!]
\begin{center}
\includegraphics[width=12cm]{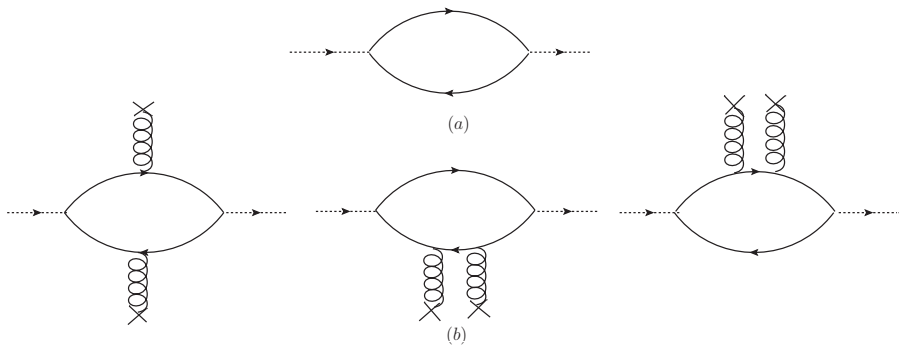}
\end{center}
\caption{ (a): Bare loop diagram
 (b): Diagrams corresponding to  gluon condensates.} \label{fig1}
\end{figure}

 In QCD side, the perturbative (short distance) and non-perturbative (long distance) parts of the correlation function  must be
 separated. In this regard, the OPE method is used and  the correlation function is  calculated in deep Euclidean region, $p^2\ll-\Lambda_{QCD}^2$. So, the correlation function is written as:
\begin{eqnarray}\label{correl.func.QCD1}
\Pi^{QCD} =\Pi_{pert}+\Pi_{nonpert}
\end{eqnarray}
The short distance contribution  (bare loop diagram  in figure
(\ref{fig1}) part (a)) is  calculated using the perturbation theory,
 whereas the long distance   contributions (diagrams shown in figure
(\ref{fig1}) part  (b) ) are parametrized in terms of   gluon condensates. To proceed,  we write the perturbative part  in terms of a dispersion integral as follows:
\begin{eqnarray}\label{correl.func.QCD1}
\Pi^{QCD} =\int \frac{ds \rho(s)}{s-p^2}+\Pi_{nonpert},
\end{eqnarray} 
where, $\rho(s)$  is  the spectral density.  This spectral density can be calculated  having the  Feynman amplitude of the bare loop diagram and using   the Cutkosky  rules. Note that, in the Cutkosky rules the quark propagators are replaced by Dirac delta function, i.e.,  $\frac{1}{p^2-m^2}\rightarrow (- 2\pi i) \delta(p^2-m^2)$.

Finally, we find the spectral
density as:
\begin{equation}
\rho(s)=\frac{3}{8\pi^2}(1-\frac{4m_b^2}{s})^{3/2}
\end{equation}
The non-perturbative part is calculated via the gluon condensate diagrams represented in part (b) of
figure (\ref{fig1}). The Fock-Schwinger
gauge,  $x^{\mu}A^{a}_{\mu}(x)=0$, is chosen. Then, the  vacuum gluon field is  as:
\begin{eqnarray}\label{Amu}
A^{a}_{\mu}(k')=-\frac{i}{2}(2 \pi)^4 G^{a}_{\rho
\mu}(0)\frac{\partial} {\partial k'_{\rho}}\delta^{(4)}(k'),
\end{eqnarray}
where $k'$ is the gluon momentum and the  quark-gluon-quark vertex is used as:
\begin{eqnarray}\label{qgqver}
\Gamma_{ij\mu}^a=ig\gamma_\mu
\left(\frac{\lambda^{a}}{2}\right)_{ij},
\end{eqnarray}
As a result, the non-perturbative part in momentum space is obtained as:

\begin{eqnarray}
\Pi_{nonpert}^i=\int^1_0\langle \alpha_s
G^2\rangle\frac{\Theta}{96\pi(m_b^2 - p^2 x + p^2 x^2)^4}dx.
\end{eqnarray}
The explicit expressions for $\Theta$ is  as follows:
  \begin{eqnarray}
\Theta &&=-\frac{1}{2} x^2 \Bigg\{2 m_b^4 x^3
   (m_b^2 (4-3 x)+p^2
   (x^2+x-2))\nonumber\\ &&+m_b^4 x
   (m_b^2 (x (17-2 x (18 (x-3)
   x+47))+8)+p^2 x (x (27 x-25)-7)
   (x-1)^2)\nonumber\\ &&+m_b^2 (-m_b^2 p^2 (x-1) x
   (x (x (2 x (27 x-82)+149)-32)-11)\nonumber\\ &&+m_b^4 (3 x
   (x (x (3 x (4 x-17)+76)-46)+8)+5)+p^4 (x-1)^3
   x^2 (24 x^2-22 x-5))\nonumber\\ &&+m_b^2 (x-1) x^2 (m_b^2 p^2 (7-x (2 x+3))+m_b^4
   (3 x-5)\nonumber\\ &&+p^4 (x-1) (2 (x-1) x+3))+(x-1)
   (-m_b^2 p^4 (x-1) x (2 x (x (2 x (6
   x-19)\nonumber\\ &&+37)-10)-3)+m_b^4 p^2 (x (x (x (x (27
   x-112)+162)-97)+14)+5)\nonumber\\ &&+m_b^6 (x (57-2 x (3 x
   (2 x-9)+43))-15)+p^6 (x-1)^3 x^2 (6 (x-1)
   x-1))\nonumber\\ &&+3  m_b^6 x^4+3 m_b^6 (x-1)^2
   x^2 (4 x+1)\Bigg\}.
\end{eqnarray}

 The next step is to apply Borel transformation over  $p^2$ for suppressing  the contribution of the higher
states and continuum. Then, sum rules are obtained as:
\begin{eqnarray}\label{lepsum}
m_{A}^2f_{A}^2e^{\frac{-m_{A}^2}{M^2}}&=&
\int_{4m_b^2}^{s_{0}}
ds~\rho(s)~e^{-\frac{s}{M^{2}}}+\hat{B}\Pi_{nonpert},
\end{eqnarray}
where $M^2$ and $s_{0}$ are the Borel mass parameter and the continuum threshold, respectively. 

The meson mass $(m_{A})$ is extracted with simple algebra (applying derivative with respect to $-\frac{1}{M^2}$  to the both
sides of the above  sum rule  and dividing by itself). It is obtained as follows:
\begin{eqnarray}\label{mass2}
m_A^2=\frac{-\frac{d}{d(\frac{1}{M^2})}\left[\int_{4m_b^2}^{s_{0}}
ds~\rho^{A}(s)~e^{-\frac{s}{M^{2}}}+\hat{B}\Pi^{A}_{nonpert}\right]}{\int_{4m_b^2}^{s_{0}}
ds~\rho^{A}(s)~e^{-\frac{s}{M^{2}}}+\hat{B}\Pi^{A}_{nonpert}},
\end{eqnarray}
where
\begin{eqnarray}
\hat{B}\Pi^i_{nonpert}=\int^1_0 e^{\frac{m_b^2}{M^2x(x-1)}}\frac{
 \Delta}{\pi 96 M^6(x-1)^4x^3}\langle \alpha_s G^2\rangle dx,
\end{eqnarray}
and we get
\begin{eqnarray}
\Delta &&=- m_b^4 (x-1) x^2 (m_b^2 (2 x^2-5
   x+2)\\ &&+2 M^2 x (x^2-4
   x+3))-m_b^4 (x-1) x^3 (m_b^2
   (3 x^2-6 x+2)\nonumber\\ &&+M^2 x (3 x^2-4
   x+1))+m_b^2 (x-1) x (m_b^2
   M^2 x (2 x^3-11 x^2+17 x-6)\nonumber\\ &&+m_b^4
   (x^3-4 x^2+4 x-1)+2 M^4 x^2
   (2 x^3-4 x^2+5 x-3))+m_b^2 x
   (m_b^2 M^2 x (6 x^5\nonumber\\ &&-26 x^4+43
   x^3-31 x^2+9 x-1)+m_b^4 (3 x^5-15
   x^4+27 x^3-20 x^2+7 x-1)\nonumber\\ &&+4 M^4 x^2 (3
   x+1) (x-1)^4)+(x-1) (-2 m_b^2 M^4
   x^3 (6 x^4-28 x^3+45 x^2-31
   x+8)\nonumber\\ &&+m_b^4 M^2 x (-3 x^5+16 x^4-33
   x^3+28 x^2-11 x+2)-m_b^6 (x^5-6
   x^4+13 x^3-11 x^2\nonumber\\ &&+5 x-1)+2 M^6 (x-1)^3
   x^3 (6 x^2-6 x-1))+m_b^6
   (x-1) x^5+ m_b^6 (x-1)^2 x^3.\nonumber
\end{eqnarray}

\section{Numerical Results}
The sum rules for the decay constant and the mass of the  $h_b(1P)$ depend on the bottom quark mass, QCD parameters and two auxiliary parameters  namely Borel mass parameter $M^2$ and continuum
threshold $s_0$.  The values $m_b=4.8~GeV$ and ${\langle}0\mid \frac{1}{\pi}\alpha_s G^2 \mid 0 {\rangle}=0.012~GeV^4$~\cite{Ioffe:2002be}  are chosen. Even though the physical observables suppose to be  independent of auxiliary parameters, the weak dependency of physical observables  on auxiliary parameters can be tolerated.  We have analysed the dependence of the mass and the decay constant on the Borel mass parameter $M^2$ and the continuum threshold $s_{0}$ as shown by figs. (2) and (3). With these figures, the  interval of auxiliary parameters can be fixed  in principle when the sensitivity  to the variation of these parameters are weak. Furthermore, the interval of  the Borel mass parameter is determined when the higher state and continuum contributions and the contributions of the highest order operators are
suppressed; i.e., the sum rules for the mass and the decay constant  converge. As a result of the procedure mentioned above, the
interval of the Borel parameter is taken to be $ 15~ GeV^2\leq M^2 \leq 30~ GeV^2 $. Note that the similar region has been found for analysing the mass and the decay constant for ${\cal{X}}_{b0}$~\cite{Veliev:2010gb}.
The continuum threshold $s_{0}$ is not completely arbitrary  but it is  related to
the energy of the first exited states ($h_b(2P)$) with the same quantum numbers as the interpolating currents. In other world, the interval of $s_0$ must be in between  the mass square  of ($h_b(1P)$) and ($h_b(2P)$), i.e., $m^2[h_b(1P)]<s_0<m^2[h_b(2P)]$,  where $m[h_b(1P)] = (9898.25\pm1.06^{+1.03}_{-1.07} )~\textsl{MeV}$ and $m[h_b(2P)] = (10259.76\pm1.064^{+1.43}_{-1.03} )~\textsl{MeV}$. Then, the interval
$101~GeV^2\leq s_0\leq 103~GeV^2$ for the continuum threshold $s_0$ is chosen. Here, it is worth to be mentioned that in the above regions for Borel mass parameter $M^2$ and continuum threshold $s_{0}$ the dependence of the results on auxiliary parameters is the most weak one. 

Regarding the intervals of the Borel mass parameter $M^2$, continuum threshold $s_0$ and input parameters mentioned above, the following values for the mass and the decay constant of $h_b(1P)$ are obtained:
\begin{eqnarray}
m[h_b(1P)]&&=(9940\pm 37)~\textsl{MeV}\\
f[h_b(1P)]&&=(94\pm 10)~\textsl{MeV}
\end{eqnarray} 
Note that, we use the experimental value of the mass $m[h_b(1P)] = 9898.25~\textsl{MeV}$ rather than the theoretical value of that in order to determine the decay constant. It is also worth to mention that the result for the value of the mass is in good agreement with the recent data on the mass of $h_b(1P)$ by Belle collaboration \cite{Adachi:2011ji} where they found:
\begin{eqnarray}
m[h_b(1P)] = (9898.25\pm1.06^{+1.03}_{-1.07} )~\textsl{MeV}
\end{eqnarray}

 The mass  and the decay constant of axial vector $ h_b(1P) $  are depicted in figures (\ref{fig4}-\ref{fig5}) at three  fixed values of the continuum threshold in terms of Borel mass parameter $M^2$.
These figures indicate a good stability of the mass and decay constant with respect to the Borel mass parameter $M^2$. From these figures, it is also clear that the results depend  weakly on the continuum threshold in its working region. 

In summary, this study presents  first theoretical calculation on the mass and decay constant of the newly found axial vector meson $h_b(1P)$  in the framework of the two point QCD sum rules. While the result for the  mass is consistent with the measured value by Belle collaboration, the decay constant has not yet been measured.

\begin{figure}[h!]
\begin{center}
\includegraphics[width=12cm]{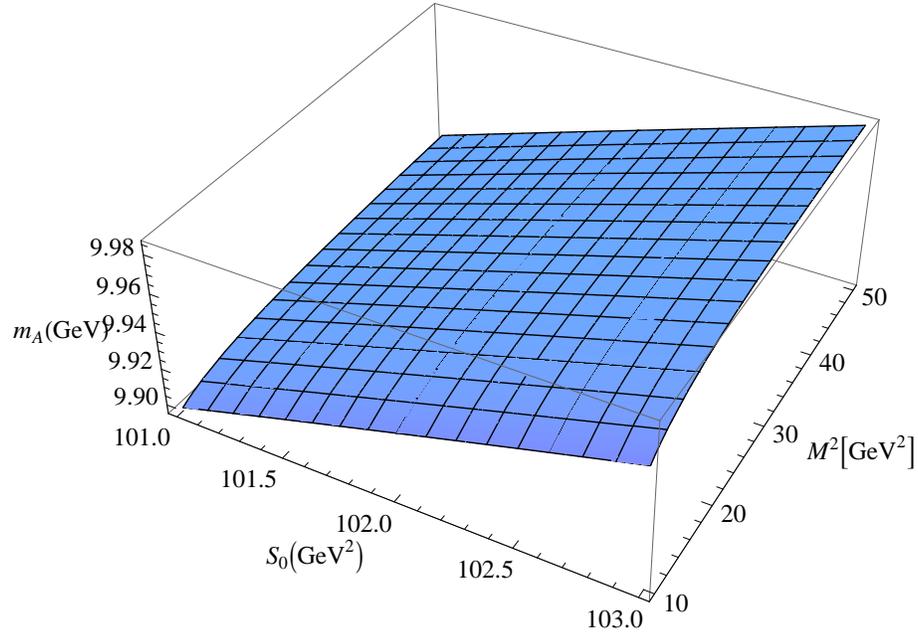}
\end{center}
\caption{Dependence of the mass of the axial vector $\bar{b}\gamma_\mu \gamma_5 b$   on the Borel parameter, $M^{2}$  and 
 the continuum threshold $s_0$. 
.} \label{fig2}
\end{figure}

\begin{figure}[h!]
\begin{center}
\includegraphics[width=12cm]{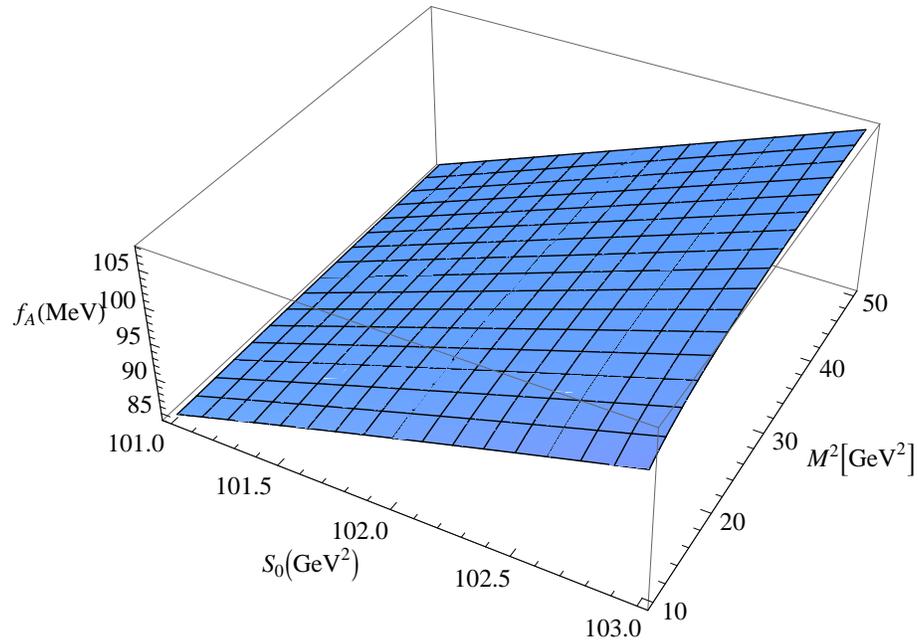}
\end{center}
\caption{Dependence of decay constant of the axial vector $\bar{b}\gamma_\mu \gamma_5 b$   on the Borel parameter, $M^{2}$  and 
 the continuum threshold $s_0$. 
.} \label{fig3}
\end{figure}

\begin{figure}[h!]
\begin{center}
\includegraphics[width=12cm]{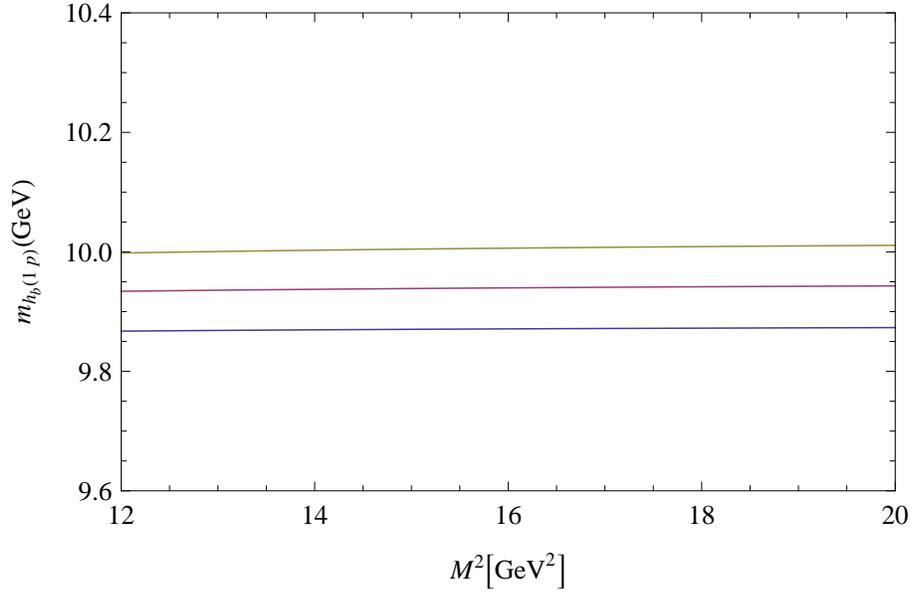}
\end{center}
\caption{Dependence of the mass of the axial vector $\bar{b}\gamma_\mu \gamma_5 b$   on the Borel parameter, $M^{2}$  at three fixed
values of the continuum threshold. The upper, middle and lower lines belong to the values $s_0=103~GeV^2$, $s_0=102~GeV^2$ and $s_0=101~GeV^2$, respectively } \label{fig4}
\end{figure}
\begin{figure}[h!]
\begin{center}
\includegraphics[width=12cm]{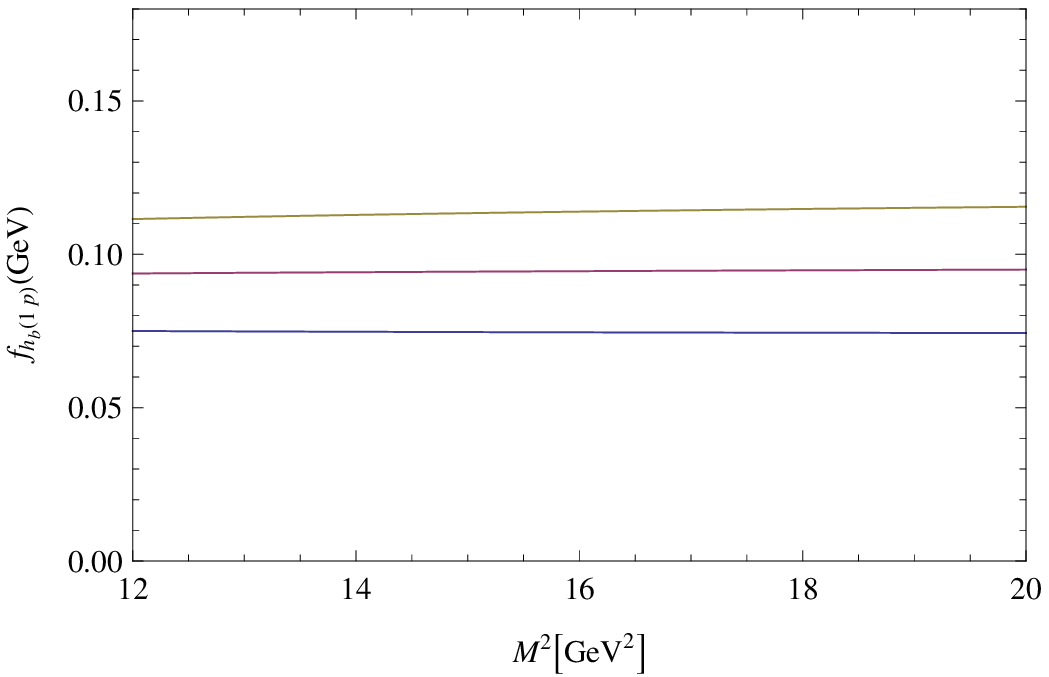}
\end{center}
\caption{Dependence of the decay constant of the axial vector  $\bar{b}b$   on the Borel parameter, $M^{2}$  at three fixed
values of the continuum threshold. The upper, middle and lower lines belong to the values $103~GeV^2$, $s_0=102~GeV^2$ and $s_0=101~GeV^2$, respectively.} \label{fig5}
\end{figure}


\begin{thebibliography}{99}
\bibitem{Adachi:2011ji}
  I.~Adachi {\it et al.}  [Belle Collaboration],
  arXiv:1103.3419 [hep-ex].


\bibitem{Brambilla}  N. Brambilla et al., Eur.Phys.J.C 71, 1534 (2011).
\bibitem{svz} M. A. Shifman, A. I. Vainshtein, V. I. Zakharov, Nucl. Phys. B 147 (1979) 385.
 \bibitem{colangelo} P. Colangelo, A. Khodjamirian, in "At the Frontier of Particle
Physics/Handbook of QCD", edited by M. Shifman (World Scientific,
Singapore, 2001), Vol. 3, p. 1495.
\bibitem{braun}  V. M. Braun, arXiv: hep-ph/9801222.
\bibitem{balitsky} I. I. Balitsky, V. M. Braun, A. V. Kolesnichenko, Nucl.
Phys. B 312 (1989) 509.

\bibitem{Kazem1} T. M. Aliev, K. Azizi, V. Bashiry, J. Phys.
G 37, 025001 (2010).
\bibitem{Bashiry:2011pp}
  V.~Bashiry, K.~Azizi and S.~Sultansoy,
  Phys.\ Rev.\  D {\bf 84}, 036006 (2011)
  [arXiv:1104.2879 [hep-ph]].
 
\bibitem{Narison}
S.~Narison, Phys.\ Lett.\ B{\bf 387}, 162 (1996). 

\bibitem{Jamin:1997rt}
  M.~Jamin, A.~Pich,
  Nucl.\ Phys.\  {\bf B507}, 334-352 (1997).
  [hep-ph/9702276].

  
\bibitem{Ioffe:2002be}
  B.~L.~Ioffe, K.~N.~Zyablyuk,
  Eur.\ Phys.\ J.\  {\bf C27}, 229-241 (2003).
  [arXiv:hep-ph/0207183 [hep-ph]].

\bibitem{Veliev:2010gb}
  E.~V.~Veliev, H.~Sundu, K.~Azizi, M.~Bayar,
  Phys.\ Rev.\  {\bf D82}, 056012 (2010).
  [arXiv:1003.0119 [hep-ph]].

   \end{thebibliography}
\end{document}